\documentclass{emulateapj}
\usepackage{amssymb, amsmath}

\newcommand{\kel}{\mbox{ K}}

\newcommand{\Mpc}{\mbox{ Mpc}}

\newcommand{\hunits}{\mbox{ km s$^{-1}$ Mpc$^{-1}$}}
\newcommand{\bq}{\begin{equation}}
\newcommand{\eq}{\end{equation}}
\newcommand{\bqa}{\begin{eqnarray}}
\newcommand{\eqa}{\end{eqnarray}}

\newcommand{\xh}{x_H}

\newcommand{\qperp}{{\bf q}_{\perp}({\bf k})}
\newcommand{\qperpp}{{\bf q}^*_{\perp}({\bf k}')}
\newcommand{\bfk}{{\bf k}}
\newcommand{\bfx}{{\bf x}}
\newcommand{\Pxx}{P_{\delta_{x} \delta_{x}}}
\newcommand{\Pxd}{P_{\delta_{x} \delta_{b}}}

\newcommand{\hk}{{\bf \hat{k}}}
\newcommand{\Cl}{{\it l}^2 C_{\it l}/ {2 \pi}}
\newcommand{\deriv}{{\rm d}}

\begin{document}

\title{The Kinetic Sunyaev-Zel'dovich Effect from Reionization}

\author{Matthew McQuinn,\altaffilmark{1} Steven
R. Furlanetto,\altaffilmark{2} Lars Hernquist,\altaffilmark{1} Oliver
Zahn,\altaffilmark{1} \& Matias Zaldarriaga\altaffilmark{1,3} }

\altaffiltext{1} {Harvard-Smithsonian Center for Astrophysics, 60
Garden St., Cambridge, MA 02138; mmcquinn@cfa.harvard.edu}

\altaffiltext{2} {Division of Physics, Mathematics, \& Astronomy;
  California Institute of Technology;  Pasadena, CA
  91125}

\altaffiltext{3} {Jefferson Laboratory of Physics, Harvard University,
Cambridge, MA 02138}

\begin{abstract}
During the epoch of reionization, local variations in the ionized
fraction (patchiness) imprint arcminute-scale temperature anisotropies
in the CMB through the kinetic Sunyaev-Zel'dovich (kSZ) effect.  We
employ an improved version of an analytic model of reionization
devised in \citet{furlanetto04a} to calculate the kSZ anisotropy from
patchy reionization.  This model uses extended Press-Schechter theory
to determine the distribution and evolution of \ion{H}{2} bubbles and
produces qualitatively similar reionization histories to those seen in
recent numerical simulations.  We find that the angular power spectrum
of the kSZ anisotropies depends strongly on the size distribution of
the \ion{H}{2} bubbles and on the duration of reionization.  In
addition, we show that upcoming measurements of the kSZ effect should
be able to distinguish between several popular reionization scenarios.
In particular, the amplitude of the patchy power spectrum for
reionization scenarios in which the IGM is significantly ionized by
Population III stars (or by mini-quasars/decaying particles) can be
larger (or smaller) by over a factor of 3 than the amplitude in more
traditional reionization histories (with temperature anisotropies that
range between $0.5$ and $3 ~ \mu K$ at ${\it l} = 5000$).  We
highlight the differences in the kSZ signal between many possible
reionization morphologies and discuss the constraints that future
observations of the kSZ will place on this epoch.
\end{abstract}


\section{Introduction}
\label{intro}

Scattering of cosmic microwave background (CMB) photons off objects
after recombination imprints hot and cold spots in the CMB.
Measurement of these secondary anisotropies will elucidate details of
the formation and evolution of structure in the universe, including
the morphology of reionization. CMB detectors (such as BIMA, CBI and
soon ACT and SPT\footnote{For more information, see
http://bima.astro.umd.edu, http://www.astro.caltech.edu/$\sim$tjp/CBI,
http://www.hep.upenn.edu/act/act.html, http://astro.uchicago.edu/spt/,
respectively.}) are beginning to reach small enough scales ($\lesssim
5$ arcminutes) where the primordial temperature anisotropies no
longer dominate the secondaries.  On these scales, the principal
secondary anisotropy at most wavelengths is the thermal
Sunyaev-Zel'dovich (tSZ) effect, which comes from scattering off hot
intracluster gas \citep{zeldovich69}.  The tSZ signal is dominated by
nearby clusters and therefore is not optimal for studying the
high-redshift universe. However, the unique frequency dependence of
the tSZ anisotropies (vanishing at $\sim 217$ GHz) facilitates their
removal from the signal. This may allow future CMB missions to detect
the kinetic Sunyaev-Zel'dovich (kSZ) effect---owing to scattering off of
objects with bulk peculiar motions \citep{sunyaev80}.

During the epoch of reionization, local variations in the ionized
fraction (or patchiness) contribute to the kSZ signal.  The
anisotropies produced by this patchiness have been calculated for
various analytic models of reionization \citep{gruzinov98,
knox98,valageas01, santos03}.  These treatments differ considerably in
their description of this epoch, but most find that the kSZ
temperature anisotropies produced during reionization have a similar
amplitude to the kSZ anisotropies produced after reionization.  This
suggests that observations of the kSZ effect might provide important
constraints on the reionization epoch.

Existing probes of this era have provided few definitive answers about
reionization. Observations of Ly$\alpha$ absorption in the spectra of
high-redshift quasars indicate that it ends at $z \sim 6$
\citep{becker,fan,white03,sokasian03,wyithe04-prox,mesinger04,oh04}.
The main difficulty with these measurements is that the Ly$\alpha$
optical depth is extremely large in a neutral medium \citep{gunn65},
making it difficult to place strong constraints when the neutral
fraction exceeds $\sim 10^{-3}$.  On the other hand, measurements of
the cosmic microwave background by the \emph{Wilkinson Microwave
Anisotropy Probe} (\emph{WMAP}) suggest a high optical depth to
electron scattering, apparently requiring reionization to begin at $z
\ga 14$ \citep{kogut03,spergel03}.  Unfortunately, the \emph{WMAP}
data provide only an integral constraint on the reionization history.

In this paper, we employ the analytic model outlined in
\citet[hereafter FZH04]{furlanetto04a} to calculate the kSZ signal for
many different reionization histories.  Most analytic models of
reionization are based on the growth of \ion{H}{2} regions around
individual galaxies or quasars \citep{arons72,barkana01}.  This
contrasts with current state-of-the-art simulations
\citep{sokasian03,sokasian04,ciardi03-sim,furl04d}, which find a
relatively small number of large ionized regions around clusters of
sources.  Moreover, in these simulations, reionization proceeds from
high to low density regions, implying that recombinations play only a
secondary role in determining the morphology of reionization; instead,
large-scale bias dominates.  The model we employ produces similar
reionization histories to those seen in these simulations.

This paper is organized as follows.  We outline the formalism for
calculating the kSZ power spectrum in \S \ref{kSZ}.  In \S \ref{ps},
we review and improve upon the reionization model first developed in
FZH04.  This model is able to produce a range of possible
morphologies for reionization.  We then calculate the kSZ power
spectrum for several different reionization scenarios, highlighting
the observable differences between them (\S \ref{pr}),
and discuss whether future observations of the kSZ effect will be able to
constrain the history of reionization (\S \ref{discussion}).

In our calculations, we assume a cosmology with $\Omega_m=0.3$,
$\Omega_\Lambda=0.7$, $\Omega_b=0.046$, $H=100 h \hunits$ (with
$h=0.7$), $n=1$, and $\sigma_8=0.9$, consistent with the most recent
measurements \citep{spergel03}.

\section{The kinetic SZ effect}
\label{kSZ}

Thomson scattering of CMB photons off free electrons with a bulk
peculiar velocity produces a temperature anisotropy along the line of
sight ${\bf \hat{n}}$:
\begin{equation}
\frac{\Delta T}{T}({\bf \hat{n}}) = \sigma_T
    \; \int{ \deriv \eta \; e^{-\tau(\eta)} \; a\, n_e \;
{\bf \hat{n}} \cdot {\bf v}  },
\label{anisotropy}
\end{equation}
\noindent where $a$ is the scale factor, $\tau (\eta)$ is the Thomson
optical depth to the scatterer at conformal time $\eta$, ${\bf v}({\bf
  \hat{n}}, \eta) $ is the peculiar velocity of the scatterer, and the
electron number density is
\begin{equation}
n_e(\eta, {\bf \hat{n}}) = \bar{n}_e (\eta) \; \bar{x}_i(\eta) \; [1
+ \delta_b(\eta, {\bf \hat{n}})+ \delta_{x}(\eta, {\bf \hat{n}})].
\label{eq:ne}
\end{equation}
\noindent Here, $\bar{n}_e$ is the average number density of electrons
(both within atomic systems and free), $\bar{x}_i$ is the global ionized
fraction, and $\delta_{b}$ and $\delta_x$ are the overdensities in the
baryonic mass and the ionized fraction (we reserve $\delta$ for the
matter overdensity).\footnote{Note that eq. (\ref{eq:ne}) is not a
formally rigorous perturbation expansion.  However, it suffices for
our purposes because the second order $\delta_x \delta_{b}$ term
contributes a negligible fraction of the total kSZ signal. }  The 
two-point correlation function $w(\theta) = \langle \frac{\Delta
T}{T}({\bf \hat{n}})\frac{\Delta T}{T}({\bf \hat{n}'})\rangle $ is
\begin{eqnarray}
    w(\theta) & =& [\sigma_T \,\bar{n}_e(\eta_0)]^2 \; \int d\eta \;
    {W(\eta)} \int d\eta' \,{W(\eta')}\, \hat{n}_i\,
    \hat{n}_j' \nonumber \\
    &  & \times \langle q^i \, {q'}^j\rangle (\sqrt{\theta^2 \,
    (\eta_0 - \eta)^2 + (\eta - \eta')^2} \, ,\, \eta), \label{wtheta}
\end{eqnarray}
where ${\bf q} ({\bfx}, \eta) = {\bf v}\,(1 + \delta_b + \delta_x )$,
$W(\eta) = a^{-2}\,\bar{x}_i^2 \,e^{-\tau}$, and repeated indexes are
summed.  In equation (\ref{wtheta}), correlations between points
separated by the vector $(\Delta \eta, \Delta x)$ are treated as
equivalent to points separated by $(0, \Delta x)$.  While this
approximation should be adequate for our purposes, we substitute
$\sqrt{\langle {\bf q} \, {\bf q}'\rangle(\bfx, \eta)~\langle {\bf q}
\, {\bf q}'\rangle(\bfx, \eta')}$ for $\langle {\bf q} \, {\bf
q}'\rangle (x, \eta)$ to capture better the $\eta$ dependence.  

Simplifications to equation (\ref{wtheta}) are most
apparent in Fourier space: equation (\ref{anisotropy}) involves the
integral $\int{ d\eta \, W(\eta) \, {\bf \hat{n}} \cdot {\bf q}({\bf
k}) \, e^{{\it i} {\bf k} \cdot \hat{n}\, (\eta_0 - \eta)}}$, which
suffers severe cancellation for $\bfk$ along the line of sight.  Thus,
for modes much shorter than the thickness of the window function and
in the flat sky approximation ($k = {\it l}/x$, $x = \eta_0 - \eta$),
the Fourier space version of equation (\ref{wtheta}) reduces to
\citep{kaiser92, jaffe98}
\begin{equation}
    C_{\it l} =  (\sigma_T \,\bar{n}_e(\eta_0))^2 \; \int{\frac{d\eta}{x^2} \;
    {W(\eta)}^2 \, P_{q_\perp}({\it l}/x, \eta)},
    \label{cl}
\end{equation}
where $\langle \qperp \cdot \qperpp \rangle  = 2 \,(2 \pi)^3 \,
\delta^3_D(\bfk - \bfk')\, P_{q_\perp}(\bfk)$ and  $\qperp$ is the
projection of ${\bf q}({\bfk})$  perpendicular to ${\bfk}$,
\begin{eqnarray}
  {\bf q}_{\perp}({\bfk}) & = & \int \frac{d^3{\bfk'}}{(2\pi)^3}
   \left[\hk' - {\mu'}~ \hk \right] \, v(\bfk') \nonumber \\
  & & \times \left[\delta_b(|\bfk -
   \bfk'|) + \delta_{x}(|\bfk - \bfk'|) \right].
\label{q_perp}
\end{eqnarray}
Here, $\mu' = \hk \cdot \hk'$ and ${\bf v}({\bfk}) = v(k)\, {\bf
\hat{k}}$ where ${\bf v}({\bf k})$ is the linear theory velocity field
[${\bf v}({\bf k}) = {\it i}\,{\bf \hat{k}} \,(f \dot{a}/k)\,
\delta^{\rm lin}(k, \eta)$, with $f = d \log D/d\log a$ and $D$ the
growth factor].  The linear theory velocity suffices because nonlinear
contributions to the the velocity of objects are negligible in the
early universe.  \citet{cooray02b} estimated the nonlinear velocity
contribution to the kSZ power spectrum by utilizing virial velocity
arguments for spherically-symmetric halos.  They found that this
contributes about an order of magnitude less signal than the kSZ
effect from the linear theory velocities.  At the high redshifts
relevant to reionization, the impact of nonlinear velocities will be
even less important.

To simplify equation (\ref{cl}), note that the three-point terms in
$P_{q_\perp}$ are negligible because they involve correlations with
just ${\bf v}({\bf k})$ at one point, and ${\bf
v}_\perp(\bfk) = {\bf 0}$ by definition.  This leaves three four-point
correlation functions.  We can decompose these into a sum of all
possible pairs of two-point functions along with the connected fourth
moments \citep{ma02, santos03}
\begin{eqnarray}
P_{q_\perp}(k) & = & \frac{1}{2} \sum_{a, b =\delta_{b}, \delta_x}
\int \frac{\deriv^3{\bf k'}}{(2 \pi)^3} [(1 - {\mu'}^2) P_{vv}(k') P_{a b}(|{\bf k} - {\bf k}'|)  \nonumber \\
& &- \frac{(1 - {\mu'}^2)\,
k'}{|{\bf k} - {\bf k}'|} P_{av}(k') P_{b v}(|{\bf k} - {\bf k}'|) ] \nonumber \\
& & + \int \frac{\deriv \bfk' ~ \deriv
\bfk''}{(2 \pi)^6} \,\sqrt{(1 - {\mu'}^2)\,(1 - {\mu''}^2)}\, {\rm
cos}(\phi' -\phi'') \nonumber \\
& & \times P_{\delta_a \delta_b v v}(\bfk - \bfk', -\bfk
-\bfk'', \bfk', \bfk''), 
\label{P_q}
\end{eqnarray}
\noindent where $P_{\delta_* \delta_* v v}$ is the connected fourth
moment and $\{\phi', \phi''\}$ are the polar angles of the vectors
$\{\bfk', \bfk''\}$.  In the halo model and in our reionization model,
where \ion{H}{2} bubbles are assumed spherical and correlations
between objects are handled similarly to those in the halo model,
$P_{\delta_* \delta_* v v}$ does not depend on $\phi'$.  Therefore the
contribution from the connected fourth moments vanish.  \citet{ma02}
verify in simulations that the contribution to $P_{q_\perp}$ from
$P_{\delta_b \delta_b v v}$ is indeed negligible.  \citet{zahn05}
compute the kSZ for a nearly identical distribution of bubbles sizes
as we do, but allow for more complicated configurations of bubbles.
The agreement of our kSZ signal with theirs (see
Fig.~\ref{combinedfig} and \S \ref{sr} below), suggests that the
connected fourth moments involving $\delta_x$ are also small and
that this is not just a byproduct of our simplified model for
reionization.

To further simplify equation (\ref{P_q}), we drop the cross
correlation terms $P_{\delta_b v}$ and $P_{\delta_x v}$.  Even if
$P_{\delta_* \delta_*}\, P_{vv}$ is comparable to $P_{\delta_* v}\,
P_{\delta_* v}$, equation (\ref{P_q}) is dominated by $k'$ near zero
at the scales of interest, $k \gtrsim 1 \, {\rm Mpc}^{-1}$ (noting
that $P_{\delta_{\rm lin} \delta_{\rm lin}}$ and $P_{\delta_*
\delta_{\rm lin}}$ scale roughly as $k'^{-3}$ for $k' \gtrsim 0.05 \,
{\rm Mpc}^{-1}$ and that $v \propto \delta_{\rm lin}/k'$ ). The terms
involving $P_{\delta_* v}$ are ${\cal O}(k'/k)$ and are therefore
suppressed.

This argument suggests a further simplification for the large $k$
modes of interest.  Isolating the case $a = b = \delta_{b}$ in the sum
in equation (\ref{P_q}), dropping terms of ${\cal O}(k'/k)$, and
evaluating the angular integral, we find
\begin{equation}
P_{q_\perp,a= b = \delta_b }(k) = \frac{1}{3}\, v_{rms}^2 \,
P_{\delta_b \delta_b}(k),
\label{OV}
\end{equation}
where $v_{rms}^2 = \int{\frac{k^2 dk}{2 \pi^2} P_{vv}(k)}$
\citep{hu00, ma02}. This $P_{\delta_b \delta_b}$ contribution to
$P_{q_\perp}$ is the dominant term for uniform reionization and is
sometimes called the Ostriker-Vishniac effect (OV).  Traditionally,
this effect is calculated using linear theory \citep{ostriker86,
dodelson93}.  \citet{hu00} first calculated the nonlinear OV effect,
demonstrating that at ${\it l}\sim 10^4$ the signal is significantly
increased by nonlinearities.  We use the halo model \citep{cooray02}
to construct $P_{\delta \delta}$ -- which we then employ to calculate
the term $P_{\delta_b \delta_b}$ in equation (\ref{OV}).
Unfortunately, calculating $P_{\delta_b \delta_b}$ also requires
knowledge of the distribution of baryons inside halos: at large
overdensities, baryonic matter does not trace the dark matter.  To
model the baryon distribution, we employ a filter function $F(x)$ such
that $P_{\delta_{b} \delta_{b}}(k) = F(k) P_{\delta \delta}(k)$, where
$P_{\delta \delta}$ is the nonlinear matter power spectrum.  The
function $F(k)$ filters large $k$ contributions from the dark matter
power spectrum, such that the baryonic matter is not as clumpy as the
dark matter (see Appendix \ref{pbb} for details).

\citet{hu00} uses the same method to calculate the OV.  Our
OV signal agrees well with the analytic calculation of \citet{zhang04},
which uses a similar method but also includes the nonlinear velocity
field.  \citet{ma02} and \citet{cooray02} employ a different method to
calculate the OV effect, modeling the gas distribution in halos with a
$\beta$ profile $\rho(r) = A\,[1 + (r/r_c)^2]^{-3\beta/2}$ or a similar
function.  Their calculation results in a slightly smaller signal.
Perhaps a superior method for calculating the OV is through
hydrodynamic simulations.  Such simulations must cover a huge dynamic
range, resolving nearby halos and large scale structure at high
redshifts.  Simulations tend to agree relatively well with the
predictions of these analytic methods at high-$l$ \citep{dasilva01,
white02, zahn05}, but they predict less signal at $l < 5000$.  Such
deviations are less important for studying secondary anisotropies
because the primordial anisotropies dominate at these
scales. (Contamination of the primordial anisotropies and the
resulting bias in cosmological parameter estimation is a different
matter.)

The other two terms contributing to $P_{q_\perp}$ involve integrals
over $P_{vv}\, P_{\delta_x \delta_x}$ -- the term normally included in
calculations of patchy reionization -- and the cross correlation
$2 P_{vv} \,P_{\delta_b \delta_x}$ -- ignored in existing calculations,
but possibly significant if the \ion{H}{2} bubbles occur in overdense
regions as simulations suggest.  We refer to these two terms as the
``patchy'' terms.  Since the universe reionizes quickly in some of
our models (see Figure \ref{qfig}), the assumption implicit in
equation (\ref{cl}) -- $k^{-1}$ is much shorter than the distance over
which the window function changes appreciably -- may break down at
some relevant {\it l}.  This is not a concern for the OV effect, where
only a small portion of the signal comes from the reionization epoch.
Simple estimates suggest that at scales where the kSZ effect dominates
over the primordial anisotropies (${\it l} \gtrsim 5000$), equation
(\ref{cl}) should be an adequate approximation.  However, to be safe,
we do the calculation in configuration space (eq. \ref{wtheta}).
Additional details concerning the configuration space calculation are
given in Appendix \ref{realspace}.

\citet{hu00} showed that the high-{\it l} polarization resulting from
scattering of the primordial and kinetic quadrupoles off perturbations
in the density and ionized fraction have negligible amplitude
($10^{-3}-10^{-2} \mu K$).  This results from the small magnitude of
the quadrupole anisotropies.  The high-{\it l} polarization signal
from reionization is thus too small to affect upcoming measurements of
B-parity polarization from gravitational lensing or from gravitational
waves.  Therefore, we only discuss the temperature anisotropies here.

\section{\ion{H}{2} regions during reionization}
\label{ps}

Recent numerical simulations (e.g.,
\citealt{sokasian03,sokasian04,ciardi03-sim}) show that reionization
proceeds ``inside-out'' from high density clusters of sources to
voids, at least when the sources resemble star-forming galaxies (e.g.,
Springel \& Hernquist 2003; Hernquist \& Springel 2003).  We therefore
associate \ion{H}{2} regions with large-scale overdensities.  We
assume that a galaxy of mass $m_{\rm gal}$ can ionize a mass $\zeta
m_{\rm gal}$, where $\zeta$ is a constant that depends on: the
efficiency of ionizing photon production, the escape fraction, the
star formation efficiency, and the number of recombinations.  Values
of $\zeta \la 40$ are reasonable for normal star formation, but
very massive stars can increase the efficiency by an order of
magnitude \citep{bromm-vms}.  The criterion for a region to be ionized
by the galaxies contained inside it is then $f_{\rm coll} >
\zeta^{-1}$, where $f_{\rm coll}$ is the fraction of mass bound to
halos above some $m_{\rm min}$.  We will normally assume that this
minimum mass corresponds to a virial temperature of $10^4 \kel$, at
which point hydrogen line cooling becomes efficient.  In the extended
Press-Schechter model \citep{bond91,lacey} the collapse fraction of
halos above the critical mass $m_{\rm min}$ in a region of mean
overdensity $\delta_m$ is
\begin{equation}
f_{\rm coll} = {\rm erfc} \left( \frac{\delta_c - \delta_m}{\sqrt{2 \left[ \, \sigma^2_{\rm min} - \sigma^2(m, z) \right]}} \right),
\end{equation}
where $\sigma^2(m, z)$ is the variance of density fluctuations on the
scale $m$, $\sigma^2_{\rm min} \equiv \sigma^2(m_{\rm min}, z)$ and
$\delta_c \approx 1.686$, the critical density for collapse.  With this
equation for the collapse fraction, we can write a condition on the
mean overdensity within an ionized region of mass $m$, 
\begin{equation}
 \delta_m \ge
\delta_B(m,z) \equiv \delta_c - \sqrt{2} K(\zeta) [\sigma^2_{\rm min}
- \sigma^2(m, z)]^{1/2},
\label{eq:deltax}
\end{equation}
where $K(\zeta) = {\rm erf}^{-1}\left(1 - \zeta^{-1} \right)$.

FZH04 showed how to construct the mass function of \ion{H}{2} regions
from $\delta_B$ in an analogous way to the halo mass function
\citep{press,bond91}.  The barrier in equation (\ref{eq:deltax}) is
well approximated by a linear function in $\sigma^2$, $\delta_B
\approx B(m) = B_0 + B_1 \sigma^2(m)$, where the redshift dependence
is implicit. In that case, the mass function has an analytic expression
\citep{sheth98}: 
\bq 
n(m) = \sqrt{\frac{2}{\pi}} \
\frac{\bar{\rho}}{m^2} \ \left| \frac{\deriv \ln \sigma}{\deriv \ln m}
\right| \ \frac{B_0}{\sigma(m)} \exp \left[ - \frac{B^2(m)}{2
\sigma^2(m)} \right],
\label{eq:dndm}
\eq where $\bar\rho$ is the mean density of the universe.  Equation
(\ref{eq:dndm}) gives the comoving number density of \ion{H}{2}
regions with masses in the range $(m,m+dm)$. This result is rederived
in Appendix \ref{massfunc}.  The crucial difference between this
formula and the standard Press-Schechter mass function occurs because
$\delta_B$ is a (decreasing) function of $m$. The barrier is more
difficult to cross as one goes to smaller scales, which gives the
bubbles a characteristic size that depends primarily on $\bar{x}_i$.
The one point function $ \bar{x}_i (z)$ for the linear barrier is
(Appendix \ref{massfunc})
\begin{eqnarray}
Q(z) \equiv \bar{x}_i(z)  & = & \frac{1}{2} \, e^{-2\, B_0 \, B_1} \, {\rm
erfc} \left( \frac{B_0 - B_1 \sigma_{\rm min}^2}{\sqrt{2 \, \sigma_{\rm
min}^2}} \right) \nonumber \\
& & + \, \frac{1}{2}\, {\rm erfc} \left(\frac{B_0 + B_1 \sigma_{\rm
min}^2}{\sqrt{2 \, \sigma_{\rm min}^2}} \right). 
\label{onepoint_text}
\end{eqnarray}
This equation agrees extremely well with the $Q(z)$ one finds with the
full barrier (equation \ref{eq:deltax}).

All the quantities we want to calculate are well-defined in this
model.  In addition, since the typical bubble size is usually larger
than the scale of nonlinearities, calculating the desired patchy
correlation functions is simply a matter of exploring the properties
of a Gaussian random field.

\subsection{Bias of the \ion{H}{2} Regions}
\label{bias}

Unfortunately, there was an inaccuracy in the formula FZH04 used for the
linear bias (their eq. 22) that caused them to underestimate the
bias.  (Note that the expression in \citealt{sheth02} is also
incorrect.)  To compute the correct bias, we consider how the bubble
number density $n(m)$ varies with the underlying (large-scale)
density.  This is straightforward for a linear barrier, because the
barrier remains linear after a translation by the origin.

Suppose we are in a region of linear-extrapolated matter overdensity
$\delta_R$ and variance $\sigma_R$ at a smoothing scale $R$; note that
the linear-extrapolated matter overdensity at redshift $z$ is
$\delta_R \, D(z)$ (we adopt this notation for $\delta$ only for this
section).  The comoving number density of bubbles $n(m | \delta_R,
\sigma_R)$ is simply the usual expression, equation (\ref{eq:dndm}),
with the replacements
\begin{eqnarray}
\sigma(m)^2   &   \rightarrow   &   \sigma(m)^2 - {\sigma_R}^2,  \nonumber \\
B(m,z)     &   \rightarrow   &   B(m, z) - \delta_R, \nonumber \\
B_0        &   \rightarrow   &   B_0 + B_1 \, {\sigma_R}^2 - \delta_R. \nonumber
\label{substitutions}
\end{eqnarray}
Linearizing the resulting expression and noting that the Eulerian
overdense region is larger by a factor of $(1 + \delta_R)$
\citep{mo96} , we find
\begin{eqnarray}
\delta_x(m) & =& \delta_R \, D(z) \, \left[ 1 + \frac{B(m)/\sigma(m)^2 -
    1/B_0}{D(z)} \right] \nonumber \\
& & + \frac{B_1 \, \sigma_R^2}{B_0},
\end{eqnarray}
where $\delta_x(m)$ is the overdensity of bubbles with mass $m$ [note
that $\delta_x = \int dm \, n(m) \, \delta_x(m)$].  The last term is
only significant if $R$ is small or if we are so close to reionization
that $B_1 \sigma_R^2 \sim B_0$.  If we neglect this term, the linear
bias is thus
\begin{equation}
b_x(m) = 1 + \frac{B(m)/\sigma^2(m) - 1/B_0}{D(z)}.
\label{newbias}
\end{equation}
Interestingly, for sufficiently small bubbles $b_x < 0$.  This means
that small bubbles become rare in large overdensities and more common
in underdense regions.  Physically, this occurs because overdense
regions are more advanced in the reionization process such that the small
bubbles have already merged with larger \ion{H}{2} regions.  During
the late stages of reionization, only the deepest voids contain
galaxies isolated enough to create small bubbles.

Finally, we define the average bias 
\bq 
\bar{b}_x \equiv Q^{-1} \int dm \, n(m) \,V(m) \, b_x(m) 
\label{eq:bbar},
\eq 
which is the bias averaged over the bubble filling factor.  The
average bias tends to be quite large throughout reionization,
typically between $3< \bar{b}_x < 10$. The (incorrect) prescription of
FZH04 resulted in moderate bias $b_x \sim 1$ and thus underestimated
the clustering of bubbles.

Working only to linear order in the expansion of $\delta_R$ can be
problematic because the bias is so large.  This is
particularly true for $Q$ near unity.  In this limit, even small
overdensities can lead to the nonsensical result $x_\delta \approx Q
\, [1 +\bar{b}_x \, \delta_R \, D(z)] >1$, where $x_\delta$ is the ionized
fraction in a region of overdensity $\delta_R$.  For large $Q$, only
the most underdense parts of the universe are not ionized, so $x_\delta$
cannot be well-approximated as a linear function in $\delta_R$; the
linear bias tends to overpredict the clustering of bubbles.  
Conceptually, the problem is the same as those described in \S 3.1 of
FZH04.  The linear bias works much better for small $Q$.  Fortunately,
for large $Q$ the bubbles become so large that they are essentially
uncorrelated and the bias is unimportant.

\subsection{Autocorrelation of the Ionized Fraction}
\label{model-xixx}

In the original derivation of $\langle x_i \, x_i' \rangle$ and
$\langle x_i \, \delta'_b \rangle$, FZH04 used a Poisson model that
allowed the bubbles to overlap.  However, our expression for $n(m)$
(eq. \ref{eq:dndm}) does not actually allow this to happen: the
bubbles are explicitly constructed to be the largest contiguous
ionized regions in the universe.  Here, we construct improved versions
of $\langle x_i \, x_i' \rangle$ where overlap is forbidden.

Unfortunately, this poses a new set of problems.  To compute the
relevant integrals analytically, we must assume that the bubbles are
spherical.  This assumption tends to suppress the ionized fraction at
distances around the average bubble radius. The reason is simple.
Imagine that all the bubbles are the same size.  Then reionization
would be similar to packing oranges in a crate: there is no way to do
so without leaving some gaps, because the oranges cannot overlap.  In
reality, the bubbles will not be perfectly spherical; some will be
elongated and able to fill these gaps.  The most reliable treatment is
far from obvious.  If we completely disallow overlap, the suppression
of power at the characteristic radius of the bubbles causes
significant (and spurious) ringing in the power spectrum.  The other
option is to allow for some amount of bubble overlap.  Qualitatively,
one could think of this as allowing the bubbles to deform and fill the
gaps.  However, this will overestimate the signal at scales near the
bubble radii.  Of course this problem becomes worse as $Q \rightarrow
1$ and overlap becomes more common.

\begin{figure}
\plotone{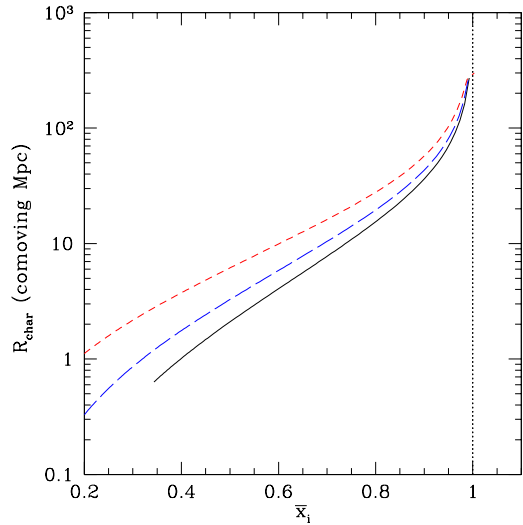} \caption{ Characteristic
bubble radius (the peak of $V dn/d{\rm log}R$) as a function of
ionized fraction $\bar{x_i}(\zeta)$ for $z = 6$ (solid), $z = 9$
(long-dashed), and $z = 15$ (short-dashed).  The \ion{H}{2} regions
become quite large for $Q \gtrsim 0.5$, which justifies our approximation
of uncorrelated bubbles at large $Q$.  All of the reionization models
considered in this paper are reionized well before $z = 6$, so the
upper two curves are most relevant.}
\label{bubble_size}
\end{figure}

Our solution is to allow bubble overlap only when the ionized fraction
is small.  Outside of this regime, we note that the characteristic
bubble radius is $\gtrsim 5$ Mpc when overlap is significant at $Q(z)
\gtrsim 0.5$.  Figure \ref{bubble_size} shows this explicitly: we plot
the characteristic bubble radius (defined as the peak of $V dn/d{\rm
log}R$) as a function of the ionized fraction at several fixed
redshifts.  When the bubbles become large, the dark matter correlation
function $\xi_{\delta \delta} (r)$ is so small on the relevant scales
that we can ignore the correlations between bubbles regardless of the
bias.  Therefore, we break our solution into two regimes: in the
first, the bubbles are small and both the one and two bubble terms,
$P_1$ and $P_2$, are important.\footnote{In analogy with the halo
model, the one and two bubble terms are respectively the contributions
to the correlation functions that arise when the points separated by a
distance $r$ are inside the same bubble ($P_1$) and are inside
different bubbles ($P_2$).} In the other, the bubbles are large and
only the one bubble term $P_1$ is significant.  We have
\begin{eqnarray}
\langle x_i x_i \rangle (r) & = \left \{ \begin{array}{l r}
(1-Q) \,P_1(r) + Q^2 \;\;\;\;\;\;\; Q > \eta, \\
P_1(r) + P_2(r) \;\;\;\;\;\;\; {\rm otherwise},  \end{array} \right.
\label{xx}
\end{eqnarray}
where
\begin{eqnarray}
P_1(r) & = & \int dm \, n(m) \, V_0(m, r) \label{onebub}, \\
P_2(r) &=& \int dm_1 \, n(m_1) \int d^3{\bf r_1} \int dm_2 \, n(m_2) \nonumber \\
& & \times \, \int d^3{\bf r_2} \, [1 + \xi({\bf r_1} - {\bf r_2} | m_1, m_2)].
\label{twobub}
\end{eqnarray}
Here, $V_0(m, r)$ is the volume within a sphere of mass $m$ that can
encompass two points separated by a distance $r$.  The ansatz $ [1-Q]
\,P_1(r) + Q^2$ for the large-bubble regime is meant to approximate
the case when we can ignore correlations between bubbles; note that it
obeys the necessary limits described in FZH04.  However, our
prescription does miss the less important large scale limit, $ \langle
x_i x_i' \rangle - Q^2 \rightarrow \bar{b}_x^2 \, \langle \delta
\delta' \rangle$.  This does not affect our calculation because
$\langle \delta \delta' \rangle$ and $\langle v v' \rangle$ are so
small on these scales -- scales where the primordial anisotropies
dominate anyway.

The integrand over ${\bf r_1}$ and ${\bf r_2}$ in equation
(\ref{twobub}) depends on our scheme for overlap, which we outline
below.  The correlation function $\xi(r | m_1, m_2)$ is the excess
probability to have a bubble of mass $m_1$ a distance $r$ from a
bubble of mass $m_2$.  For simplicity, we use the mean bubble bias
$\bar{b}_x$ throughout, so $\xi(r | m_1, m_2) = \bar{b}_x^2 \,
\xi_{\delta \delta}$, and we replace $|{\bf r_1} - {\bf r_2}|$ with
${\rm max}(r, R_1 + R_2)$, where $R_1(m_1)$ and $R_2(m_2)$ are the
bubble radii.  This allows us to take $\xi_{\delta \delta}$ out of the
volume integrals.

We, therefore, are left with four integrals for the two bubble term
$P_2$: two over mass and two over volume.  The last two can be done
analytically if we assume spherical symmetry for the bubbles and
specify some condition for overlap.  The most obvious choice is to
disallow overlap, but, as mentioned, this leads to significant
ringing.  Another possibility, which we adopt here, is to enforce the
following conditions: (1) $m_1$ cannot ionize $r_2$, and $m_2$ cannot
ionize $r_1$; (2) the center of $m_2$ cannot lie inside $m_1$, but any
other part of $m_2$ {\it can} touch $m_1$, and vice versa. To ensure
that this scheme does not drastically overestimate $\langle x_i x_i'
\rangle$ near the bubble edges, we tune $\eta \sim .5$ in equation
(\ref{xx}) to optimize agreement with the ``exact'' result---the
correlation functions for the same model constructed explicitly from a
Gaussian random field in a box of side-length $100h^{-1}$ Mpc using
the method described in \citet{zahn05}.

\begin{figure}
\plotone{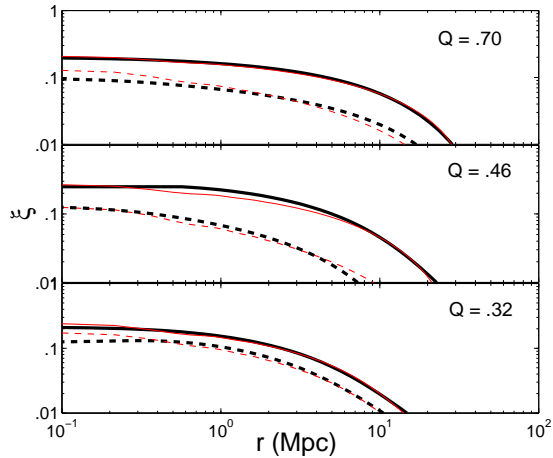} \caption{Comparison
of the analytic correlation functions in \S \ref{ps} (thick lines)
with those calculated from a Gaussian random field in a box of
side-length $100 h^{-1}$ Mpc (thin lines), using the method described
in \citet{zahn05}.  We compare these functions at three times during
reionization for $\zeta = 40$.  The solid curves are $\xi_{xx} =
\langle x_i \,x_i' \rangle - \bar{x_i}^2$ and the dashed curves are
$\xi_{x \,\delta'} = \langle x_i \, \delta_b' \rangle$.  Note that the
two methods are in good agreement.  Although, we do not expect perfect
agreement because the underlying mass function of the bubbles is
slightly different for the two methods.}
\label{justification}
\end{figure}

Figure \ref{justification} shows such a comparison for a model with
$\zeta = 40$.  Our pseudo-analytic correlation functions are in excellent
agreement with the exact result.  If we had not allowed for some
overlap or if we had set $\eta = 1$ in our prescription for $\langle
x_i ~x_i' \rangle$, the agreement would not be nearly as good.  We do not
expect perfect agreement because the exact correlation functions use a
top hat filter in real space as opposed to the $k$-space filtering
implicit in the extended Press-Schechter approach, which results in a
slightly different bubble mass function (see \S \ref{cosmology} below).

\subsection{Cross-Correlation Between the Ionized Fraction and Density}

Our prescription for $\langle x_i \, \delta_b' \rangle$ is similar to
$\langle x_i\, x_i' \rangle$.  We begin with the analog of equation
(23) in FZH04
\begin{eqnarray}
\langle x_i \delta_{b} \rangle (r) & = & -\bar{x}_i + \int dm_h \,
	\frac{m_h}{\bar{\rho}} \,  n_h(m_h) \int d^3{\bf r_h} u({\bf r} - {\bf
	r_h} | m_h) \nonumber \\
& & \times \,\int dm \, n(m) \int d^3{\bf r_b} [1 +
	\xi_{bh}({\bf r_h - r_b})],
\label{xd_init}
\end{eqnarray}
where $n_h$ is the dark matter halo number density, $u$ is the
normalized halo profile (which we can approximate as a delta function
because halos are much smaller than bubbles), and $\xi_{bh}$ is the
excess probability of having a halo of mass $m_h$ at ${\bf r}_h$ given
a bubble of mass $m$ at ${\bf r}_b$.

To evaluate equation (\ref{xd_init}), we break up the halos into those
within the same bubble as the point ${\bf r}_b$ and those outside of
it.  If ${\bf r}_h$ resides inside the same bubble of mass $m$ as
${\bf r}_b$ then we know from \S \ref{bias} that $\xi_{bh, in} =
n_h(m_h | m)/n_h(m_h) -1$.  The conditional mass function $n_h(m_h|m)$
can be computed using the extended Press-Schechter formalism
\citep{furlanetto04c}.  Thus the contribution to equation
(\ref{xd_init}) is
\begin{eqnarray}
P_{\rm in}(r) &=& \int dm \,  n(m) \, V_0(m, r) \int dm_h \frac{m_h}{\rho}
\,n_h(m_h | m) \nonumber \\ & = & \int dm \, n(m) \, V_0(m, r) \,
[1 + \delta_B].
\label{eq:pin}
\end{eqnarray}
In the last line, we use the fact that the inner integral is simply $1
+ \delta_B$, the mean overdensity of the bubble.

If ${\bf r_h}$ is outside the bubble containing ${\bf r_b}$, then we
again use the trick $\xi_{b h} \approx b_h(m_h) \, \bar{b}_x
\,\xi_{\delta \delta}(r)$ (with the mean bias inserted for
simplicity).  Thus the contribution to the integral in equation
(\ref{xd_init}) from outside the bubble is
\begin{eqnarray}
P_{\rm out}(r) & = & \bar{x}_i - \int dm \, n(m) \, V_0(m,r) 
+ \int dm \, n(m) \nonumber \\
 & & \times \,\int
d^3{\bf r_b} \,[\bar{b}_x \,\xi_{\delta \delta}(r_{\rm eff})],
\label{eq:pout}
\end{eqnarray}
where the ${\bf r}_b$ integration is over all bubbles that ionize
${\bf r}_b$ but not ${\bf r}_h$ and, for simplicity, we evaluate
$\xi_{\delta \, \delta}$ at separation $r_{\rm eff} = {\rm max}[R(m), r ]$
rather than at ${\bf r} - {\bf r}_b$.  Note
that the second term in $P_{\rm out}$ cancels the first term in
$P_{\rm in}$.  As before, the last term becomes problematic as $Q
\rightarrow 1$; here, these difficulties arise primarily from our
adoption of linear bias, which tends to overestimate the clustering in
this limit.  Fortunately, the term $P_{\rm out}$ is unimportant at
large $Q$ because the effective bubble radius is quite large.  So
external clustering can be ignored and the one bubble term dominates.
We again break our calculation into two parts:
\begin{eqnarray}
\langle x_i \, \delta_{b} \rangle (r) & = \left \{ \begin{array}{l r}
 P_{\rm in} - P_1 \;\;\;\;\;\;& Q > \eta', \\ 
P_{\rm in} + P_{\rm out} - Q
 \;\;\;\;\;\;& {\rm otherwise}, \end{array} \right.
\label{eq_xi}
\end{eqnarray}
where $P_1$ is given by equation (\ref{onebub}).  The solution $P_{\rm
in} - P_1$ is an ansatz for when correlations with the density field
inside the bubbles dominate $\langle x_i \, \delta_b' \rangle$, which
happens when the bubbles become large.  We tune $\eta'$ to get the best
agreement with the ``exact'' $\langle x_i \, \delta_b' \rangle$; this
turns out to be when $\eta' \sim .5$.  Figure \ref{justification}
compares our expression for $\langle x_i \, \delta_b' \rangle$ with
the ``exact'' result.  We see that we are in good agreement.  Because
we smooth the density field on the scale of the bubbles and do not
include the effects of non-sphericity, we should underpredict some of
the small scale power.  This underestimate appears to be very minor,
affecting smaller scales than are relevant to our kSZ calculation.

\section{The kSZ Effect from Reionization}
\label{pr}

\subsection{Single Reionization Episodes}
\label{sr}

\begin{figure}
\plotone{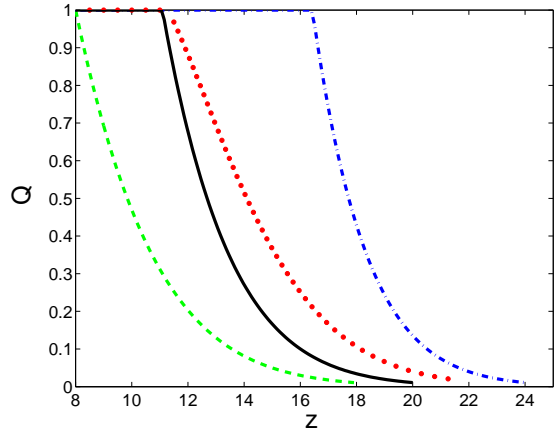}  \caption{Evolution of the
global ionized fraction for single reionization models with $\zeta =
12$, $40$, and $500$ (dashed, solid and dot-dashed, respectively). The
dotted curve assumes that $\zeta$ evolves linearly with redshift from
$\zeta = 40$ at $z \sim 11$ to $\zeta = 300$ at the start of
reionization ($z \sim 22$).  }
\label{qfig}
\end{figure}

In the simplest scenario for reionization, a single generation of
sources, such as population II stars, ionizes the universe.  At each
redshift, the ionized fraction equals the collapsed fraction times
$\zeta$ (\S \ref{ps}).  We calculate the kSZ power spectrum for models
with $\zeta$ =12, 40, and 500, which produce total optical depths of
$\tau =0.085$, $0.117$, and $0.184$, respectively (assuming that
helium is singly ionized along with hydrogen during reionization and
is fully ionized by $z = 3$; c.f. Sokasian et al. 2002).  
In addition, to
investigate the effect of a longer reionization epoch on the kSZ
signal, we calculate the power spectrum for a model where $\zeta$
increases linearly with redshift from $\zeta = 40$ when reionization
is complete ($z \sim 11$) to $\zeta = 300$ at the start of
reionization ($z \sim 22$).  This model yields an optical depth of
$\tau =0.138$.  Physically, a monotonically increasing $\zeta(z)$
could happen if lower mass halos have larger escape fractions.  The
optical depths produced by all four models are within 2-$\sigma$ of
the \emph{WMAP} best fit value of $\tau = 0.17 \pm .04$
\citep{kogut03}.  Figure \ref{qfig} plots $Q(z)$ for the four single
reionization models.

\begin{figure}
\plotone{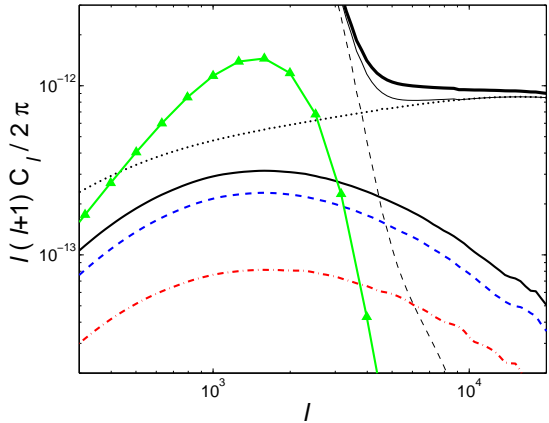}  \caption{Contributions to
the power spectrum for $\zeta = 40$. The three curves with the
smallest amplitude are the patchy contributions from $P_{\delta_x
\delta_b}$ (dot-dashed), $P_{\delta_x \delta_x}$ (dashed), and their
total (solid).  At larger amplitudes are the OV signal (dotted), the
signal from the lensed primordial anisotropies (thin dashed), and the
total power (thick solid).  In addition, we show the total signal for
a uniformly ionized medium (thin solid) and the patchy signal for the
\citet{gruzinov98} toy model with effective bubble size $R = 10$ Mpc
(solid with triangles).  Both have the same $Q(z)$ as in the $\zeta=40$ model.}
\label{i40fig}
\end{figure}

Figure \ref{i40fig} shows the various contributions to the high-{\it
l} kSZ power spectrum for the $\zeta = 40$ model.  The medium-width
solid line is the total $\Cl$ from patchy reionization -- the sum of
the contributions in equation (\ref{P_q}) that depend on $\Pxx$
(dashed) and $\Pxd$ (dot-dashed).\footnote{Note that $\Pxx(k)
\leftrightarrow \langle xx' \rangle (r)$ through a Fourier transform and
similarly for the cross correlation.}  The contribution from $\Pxx$ is
larger than the cross correlation $\Pxd$ at all scales, but the latter
is never negligible.  The dotted line is the OV power spectrum, which
is larger than the patchy signal.  The bulk of the OV effect comes
from scattering off halos at times well after reionization, and it
is therefore not optimal for studying the reionization epoch.  In
addition, we show the total high-{\it l} CMB power spectrum, comprised
of both the lensed primordial anisotropies and the kSZ contributions
(thick solid).  Patchy reionization produces a larger kSZ signal than
a uniform reionization scenario [$x_i({\bf x}, z) = \bar{x}_i(z)$]
with the same global reionization history; for comparison, we plot in
Figure \ref{i40fig} the power spectrum for uniform reionization with
the same $\bar{x}_i$ as $\zeta = 40$ (thin solid).

\citet{gruzinov98} calculated the patchy signal for a simple and
popular toy model where $\langle x_i({\bf x}) ~x_i({\bf x'}) \rangle =
e^{-({\bf x} -{\bf x'})^2/{2 \,R^2}} (\bar{x}_i - {\bar{x}_i}^2) +
{\bar{x}_i}^2$ and where the cross correlation contribution is
ignored.  The discussion in \S \ref{ps} shows that the assumption of a
Gaussian correlation function with constant standard deviation $R$
(the effective bubble size) \emph{throughout reionization} is
unrealistic.  In our model, $\langle x_i \, x'_i \rangle (r)$ is
functionally closer to a decaying exponential than a Gaussian owing to
the distribution of bubble sizes, and the shape of $\langle x_i \,
x'_i \rangle (r)$ evolves considerably during reionization as the
bubbles grow and merge.  Assuming a constant size results in a larger
peak $\Cl$ with the signal more concentrated at the specific scale.
To illustrate this point, we plot the \citet{gruzinov98} results for
$R = 10 \Mpc$ and the same ionization history as our $\zeta = 40$
model in Figure \ref{i40fig} (solid with triangles).

\begin{figure}
\plotone{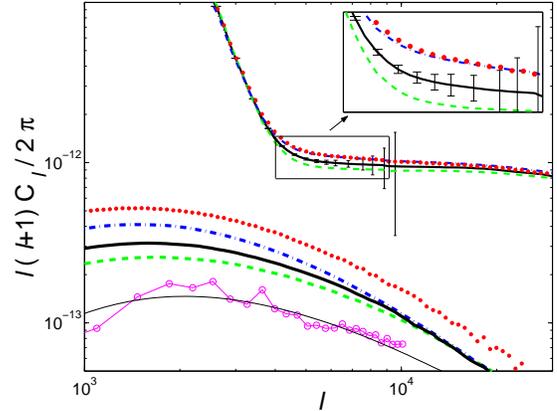}  \caption{Makeup of the
high-{\it l} signal after removal of the tSZ anisotropies.  We plot
the patchy signal (thick) and total signal (thin) for $\zeta = 12$
(dashed), $40$ (solid), and $500$ (dot-dashed), as well as a model
with variable $\zeta$ (dotted).  We include the 1-$\sigma$ error bars
for ACT assuming perfect removal of point sources and the tSZ signal.
We also plot the kSZ power from the simulation of \citet{zahn05}
(circles) and compare this to our analytic model for a comparable
$Q(z)$ (thin solid).  We do not expect perfect agreement with the
simulation because the bubble size distribution is slightly
different.}
\label{combinedfig}
\end{figure}

\begin{figure}
\plotone{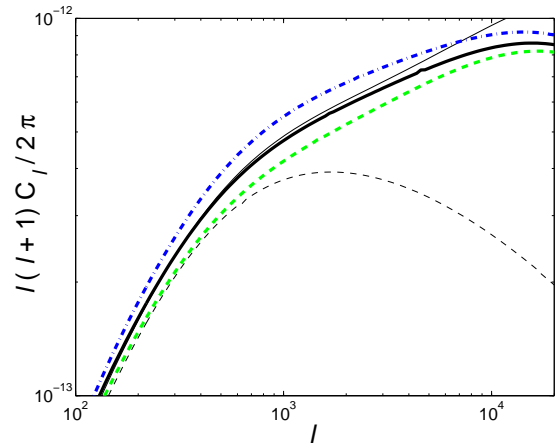} \caption{The OV contribution
to the signal for three of the single reionization models: $\zeta =
12$ (dashed), $40$ (solid), and $500$ (dot-dashed).  We also plot the
OV signal for $\zeta = 40$ in the extreme cases where the gas traces
the dark matter distribution at all redshifts (thin solid) and with the
linear theory density field (thin dashed).}
\label{OVfig}
\end{figure}

Figure \ref{combinedfig} shows the patchy signal for the four single
reionization models.  Interestingly, this power differs by less than a
factor of two between the curves.  The slight differences in the
amplitude owe mainly to the increase in density with redshift (the
probability for scattering goes as $[1+z]^{3/2}$).  This effect is
partially cancelled because the universe ionizes over a somewhat
shorter redshift interval as $\zeta$ increases.  Of the four patchy
curves in Figure \ref{combinedfig}, the evolving-$\zeta$ model has the
largest amplitude, which reflects a general theme in our results: the
amplitude of the patchy signal depends strongly on the duration of
reionization.  Finally, note that the OV contribution to the total
power is larger than that of the patchy terms (see Figures
\ref{combinedfig} and \ref{OVfig}).  However, the relative differences
between the OV contributions are comparable to the relative
differences between the patchy contributions.  Furthermore, most of
the differences between the OV signals stem from these models fully
ionizing at different times and not from differences in the duration
of the reionization epoch.

At the bottom of Figure \ref{combinedfig}, we compare the patchy
signal predicted by our analytic model to the kSZ power computed from
a $100h^{-1}$ Mpc simulation of \citet{zahn05} (their model B).  That
paper uses the same model for the patchy epoch, as outlined in \S
\ref{ps}, imposed on an SPH simulation by directly applying the
excursion set formalism to the density field of the box.  The
advantages to this approach are that it can allow for non-spherical
bubbles and includes fully nonlinear bias for the bubbles.  In
addition, the simulation should more accurately account for
nonlinearities in the density field.  The disadvantages are that
simulations encounter sample variance issues at large scales and offer
less direct physical insight into the results.  The patchy signal from
the simulation agrees well with our analytic prediction for a
comparable $Q(z)$, further justifying our simplifications in \S
\ref{kSZ} and \ref{ps}.

The ionization histories for the L20 and S5 simulations of
\citet{salvaterra05} are similar to the single reionization histories
discussed in this section.  The authors find that the functional form
and amplitude of the patchy signal for both models is similar, despite
reionization ending earlier in the L20 model.  This conclusion mirrors
what we find here: the patchy signal is mostly affected by the
duration of reionization and not by when reionization happens.  Also,
the peak amplitude of their patchy signal ($\approx 1.6 \times
10^{-13}$ for both models) is comparable to what we find.  Despite these
similarities, \citet{salvaterra05} note that their $20h^{-1}$ Mpc
simulation box may suppress the largest bubbles (particularly at the
sizes predicted by our model).  Such a bias would reduce the signal at
lower {\it l}.  It may be that simulations must employ larger boxes to
accurately predict the patchy signal.

Figure \ref{combinedfig} also plots the total signal (kSZ effect plus
lensed primordial anisotropies) for the four models (thick lines) as
well as the 1-$\sigma$ error bars for the Atacama Cosmology Telescope
(ACT) in the 225 GHz channel.  The error bars suggest that ACT will be
able to distinguish between three of the four single reionization
scenarios.  These errors were obtained from the ACT specifications of
$1.1$ arcminute resolution and $2 \ \mu {\rm K}$ sensitivity, assuming
perfect removal of point sources and the tSZ effect.  The error bars
are reliable if the frequency dependence of point source contaminants
can indeed be modeled accurately.  However, \citet{huffenberger04}
showed that there will be significant bias in the measurements of the
kSZ signal if this is not the case.  It is uncertain how well the
properties of point source contaminants can be modeled on these small
scales.

\subsection{Extended Reionization}
\label{dr}

Recently, a number of theoretical models have attempted to reconcile
the CMB and quasar data by postulating an early generation of sources
with high ionizing efficiency (most often because they contain
massive, metal free stars) along with a self-regulation mechanism that
switches to normal star formation with a lower ionizing efficiency
(e.g., \citealt{wyithe03,cen03,haiman03,sokasian03,sokasian04}). 
Such scenarios
can cause a plateau in the ionized fraction or even ``double''
reionization, in which ionized phases bracket a mostly neutral period
(although the latter possibility is unlikely;
\citealt{furl04-double}).  We refer to such scenarios as ``extended
reionization" and treat them as described in \citet{furlanetto04b}.

We model extended reionization with two generations of sources.  At
some specified redshift, the first generation of sources turns off.
For example, if the first generation consisted of massive, metal-free
stars, the natural self-regulation condition halts the formation of
these stars when the metallicity in collapsed objects passes some
threshold \citep{bromm01,bromm03,mackey03} or when hard photons
moderate H$_2$ cooling (e.g. Yoshida et al. 2003a, 2004).  Another
possibility is that photoheating halts structure formation in biased
regions.  In either case, feedback slows down the growth of \ion{H}{2}
regions by reducing the effective $\zeta$ in ionized bubbles.  We will
simplify these conditions by assuming an instantaneous transition at
some redshift $z^*$.  In reality, the transition \emph{must} be smooth
and extended \citep{furl04-double}, but for our purposes we need only
force the reionization history to stall.  Thus, in our model, the
universe develops a patchwork of \ion{H}{2} regions at $z^*$ that have
not yet overlapped completely.  The first generation imprints a set of
ionized bubbles, within which most of the second generation sources
grow (because both appear in the same overdense regions).  The bubbles
grow only slowly until the total number of ionizations from the second
generation become comparable to that of the first; after this point
the evolution approaches the normal behavior.  One important
consequence of such a treatment is that it ``freezes" a scale into the
bubbles for a long period of time \citep{furlanetto04b}.  More
realistic histories may allow the scale to vary smoothly even during
the stalling phase.

Consider a region of mass $m$ with a first generation of sources
described by $(\zeta_1,m_{{\rm min},1})$, which turn off at redshift
$z^*$, and a second generation of sources with $(\zeta_2,m_{{\rm
min},2})$.  We set $m_{{\rm min}, 2} = m_{\rm min}$ -- the mass at
which atomic cooling becomes efficient.  The total number of
ionizations is simply the sum of ionizations from the two generations.
The excursion set barrier $\delta_x(m)$ will be the solution of 
\begin{eqnarray}
1 & =& (\zeta_1 - \zeta_2) \, {\rm erfc} \left\{ \frac{\delta_c -
\delta_x(m)}{\sqrt{2[\sigma^2(m_{\rm min,1}, z') - \sigma^2(m, z')]}} \right\} \nonumber \\
& & + \, \zeta_2 \, {\rm erfc} \left\{ \frac{\delta_c -
\delta_x(m)}{\sqrt{2[\sigma^2(m_{\rm min,2}, z) - \sigma^2(m, z)]}}
\right\},
\label{eq:double}
\end{eqnarray}
where $z' = z^*$ for $z < z^*$ and $z' = z$ otherwise.  Here, the
complementary error functions are the fraction of collapsed gas above
the mass thresholds at the two redshifts.  With the new barrier, the
formalism from \S \ref{ps} carries over without further modification
(note that the barrier remains nearly linear).  Of course, our
prescription ignores the effect of recombinations.\footnote{This is
actually how we avoid double reionization even though our transition
is instantaneous; c.f. \citet{furl04-double}.}  However, even a
relatively small number of second generation ionizing sources should
be enough to halt recombinations within the \ion{H}{2} bubbles.

\begin{figure}
 \plotone{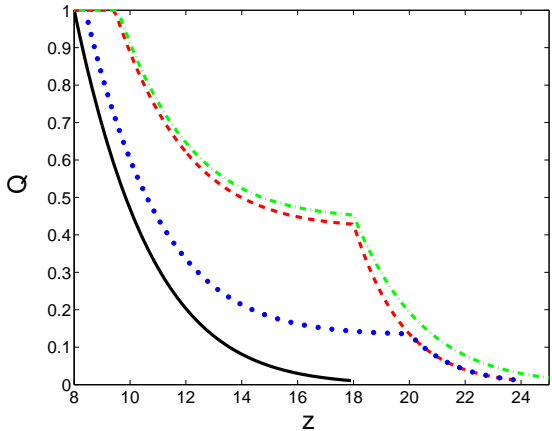}
 \caption{Evolution the global ionized fraction for
 the three extended reionization models in Table \ref{table1}.
 Model (I) is dashed,  (II) is dot-dashed,  and (III) is dotted.  We also
 plot the single reionization model with $\zeta = 12$ (solid) for
 comparison.}

\label{qdrfig}
\end{figure}

\begin{figure}
 \plotone{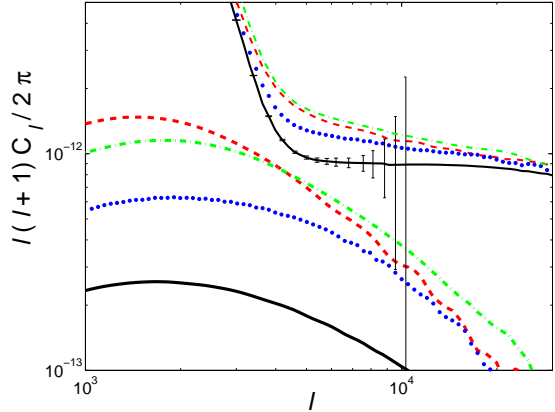}
 \caption{Thick curves show the power spectrum from patchy
 reionization and the thin curves show the total power spectrum of the
 CMB once the tSZ anisotropies are removed.  The solid curves are for
 $\zeta = 12$ and the dashed, dot-dashed, and dotted curves are for
 the extended reionization models (I), (II) and (III), respectively.  We
 include the 1-$\sigma$ ACT error bars.}
\label{drfig}
\end{figure}

Figures \ref{qdrfig} and \ref{drfig} plot the ionization histories and
angular power spectra for three extended reionization models whose
parameters and optical depths are listed in Table \ref{table1}.  The
first generation of sources turn off at $z^* = 18$ when $\bar{x}_i
\sim 0.5$ in models (I) and (II).  However, the minimum mass is ten
times smaller in model (II).  This smaller minimum mass could arise
from efficient molecular cooling within halos.  Smaller minimum masses
and smaller $\zeta_1$ both result in a somewhat smaller effective
bubble radius $R_{\rm eff}$.  Because the amplitude of the signal
scales roughly as the effective bubble size $ R_{\rm eff}$
\citep{gruzinov98}, the peak signal in model (II) is smaller than in
model (I).  However, at smaller scales, smaller bubbles cause (II) to
have slightly more power than (I).  In model (III), the first
generation of sources terminates at a high redshift, when $\bar{x}_i
\sim 0.15$.  This results in a smaller signal than the other extended
reionization scenarios but still significantly more than the models
described in \S \ref{sr}.  The OV signals for the three double
reionization models are nearly identical because the redshift of
overlap is almost the same.  As a result, the differences in the kSZ
angular power spectra stem primarily from their patchy contributions.

\begin{table}{
\caption{Parameters and optical depths for the three extended
reionization scenarios considered in this paper.  The quantity $m_{\rm
min}$ is the minimum mass at which atomic cooling becomes efficient.}

\begin{center}{
\begin{tabular}{| c c  c  c  c c |}
\hline
Model & $\zeta_1$ & $\zeta_2$ & $m_{\rm min,2}$ & $z^*$ & $\tau$\\
\hline
(I) & 12 & 500 & $m_{\rm min}$ & 18 & 0.144\\
(II) & 12 & 100 & $0.1 \times m_{\rm min}$ & 18 & 0.151 \\
(III) & 12 & 500 & $m_{\rm min}$ & 20 & 0.108\\
\hline
\end{tabular}
}\end{center}
\label{table1}
}\end{table}

All three extended reionization models result in substantially more
signal at the relevant scales than the single ionization case $\zeta =
12$ (see Fig. \ref{drfig}). This should allow future experiments to
distinguish between these two sets of scenarios relatively easily.
The reionization models considered in \citet{santos03} are most
similar to the extended scenarios considered in this section. They
find that the kSZ signal peaks around ${\it l}_{\rm peak} \sim 2000$
with an amplitude ${\it l}^2 C_l/{2 \pi} \sim 2\times 10^{-12}$.  This
${\it l}_{\rm peak}$ is at a similar location in our extended models,
and the amplitude is a bit larger than ours.  However, our signal falls off
more rapidly than theirs for ${\it l} > {\it l}_{\rm peak}$.

We note that our simplified treatment of extended reionization, with
an instantaneous transition and few recombinations, may affect some
aspects of these results.  However, the large amplitude of the patchy
component depends primarily on the duration over which the patchiness
persists and depends only weakly on our model's roughly constant
characteristic scale during this period.  Allowing the bubble size to
evolve throughout the stalling phase will tend to flatten the power
spectrum but will not substantially reduce the overall power.

\subsection{Uniform Reionization}
\label{ur}

On the scale of our bubbles, even a neutral universe is optically thin
to X-rays and gamma rays.  If sources emit a large fraction of their
energy in high energy photons
\citep{oh01b,venkatesan01,ricotti03,madau04} or if there is a particle
that decays into high energy photons at high redshifts
\citep{sciama82,hansen04}, such photons could ionize the universe more
or less uniformly.

First, consider a scenario where a decaying particle uniformly ionizes
the universe, which then recombines fully before stars turn on.  In
this case, only the OV power will be affected by the uniform
component.\footnote{Of course, the recombination process will lead to
patchiness---underdense regions will recombine later.  We expect this
patchiness to be minimal because density fluctuations are so small at
the redshifts we are interested such that a uniformly ionized IGM is a
reasonable approximation.  In principle, given a decaying particle,
calculating its effect on the kSZ signal would be a straightforward
exercise.}  However, the contribution to the OV signal from such high
redshifts is negligible anyway.  If, on the other hand, the epoch of
uniform ionization overlaps with the patchy epoch, the net effect will
be to decrease the amplitude of the patchy signal and increase the
average bubble size.  Both scenarios suggest that observations of the
kSZ effect may be able to break the optical depth degeneracy between
ionizations by a (uniform) decaying particle and discrete sources that
would imprint a patchy signal.  Alternatively, the ionizing sources
could emit both ultraviolet and high energy photons so that the
universe has both uniform and patchy components to the ionization
field.  This could occur if quasars or mini-quasars account for a
significant fraction of the ionizing photons (but see
\citealt{dijkstra04} for limits on such a scenario).

To model this possibility, we suppose that the IGM has a uniform
ionization fraction $\bar{x}_u(z)$.  On top of this lie spatial
variations from isolated \ion{H}{2} regions.  In this case, the
condition $f_{\rm coll} > \zeta^{-1}(1-\bar{x}_u)$ replaces the
barrier in our model: each galaxy can produce a larger ionized bubble
with the same number of ionizing photons.  However, rather than
varying from zero to unity, we have $0 < \xh < (1 - \bar{x}_u)$.  This
damps the fluctuations from the bubbles, requiring the rescaling
$\langle x_i x_i' \rangle - \bar{x}_i^2 \rightarrow (1-\bar{x}_u)^2
\,(\langle x_i x_i' \rangle - \bar{x}_i^2)$ and $\langle x_i \delta_b'
\rangle \rightarrow (1-\bar{x}_u) \langle x_i \delta_b' \rangle$ (this
increases the importance of the cross-correlation term).  Otherwise
our formalism is unchanged.

\begin{figure}
 \plotone{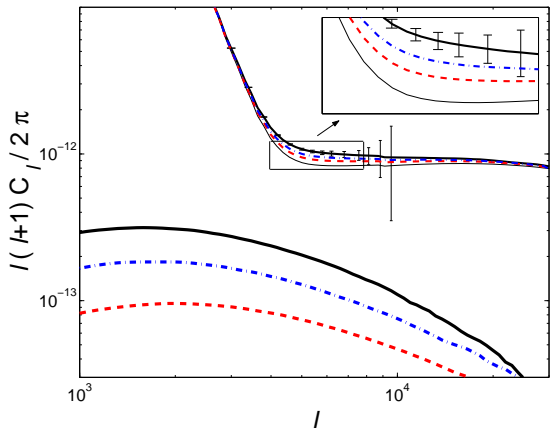}  \caption{Same as
Figure \ref{drfig}, but with curves for $\zeta = 40$ (solid) and for
three uniform reionization scenarios with $\zeta = 40 \,(1- \mu)$ and
$\bar{x}_u = 40\, \mu\, {\it f}_{\rm coll}$.  We show $\mu = 0.25$
(dot-dashed), $\mu = 0.5$ (dashed) and $\mu = 1.0$ (thin solid).
These models have the same $Q(z)$ as the canonical $\zeta = 40$ model
and therefore the same OV signal.  We include the 1-$\sigma$ ACT error
bars.}
\label{unifig}
\end{figure}

In Figure \ref{unifig}, we plot reionization scenarios where the
uniform component is proportional to the total ionized
fraction, $\bar{x}_u = \mu \,\zeta \, f_{\rm coll}$ [it follows that
the patchy fraction is $x_p = (1-\mu) \,\zeta \, f_{\rm coll}$].  We
plot results for $\zeta = 40$ with proportionality factors $\mu = 0.0$
(solid), $\mu = 0.25$ (dot-dashed), $ \mu = 0.5$ (dashed) and $\mu =
1.0$ (thin solid).  The rough interpretation is that $\mu$
parameterizes the fraction of ionizations caused by high energy
photons emitted by discrete sources.  The choice $\mu = 0.0$
corresponds to normal patchy reionization and $\mu = 1$ to purely
uniform reionization.
The ionized fraction $\bar{x}_i(z)$ is the same for all of these
curves (the solid line in Figure \ref{qfig}).  Note that the power
spectrum for $\mu = 0.25$ and $\mu = 0.5$ is suppressed by a factor of
$\sim 1.5$ and $\sim 3.0$ from the patchy model $\mu = 0.0$.  The
1-$\sigma$ error bars in Figure \ref{unifig} suggest that ACT is
capable of distinguishing between these models.

\subsection{Reionization and Uncertainties in the Underlying Cosmology}
\label{cosmology}

\begin{figure}
\plotone{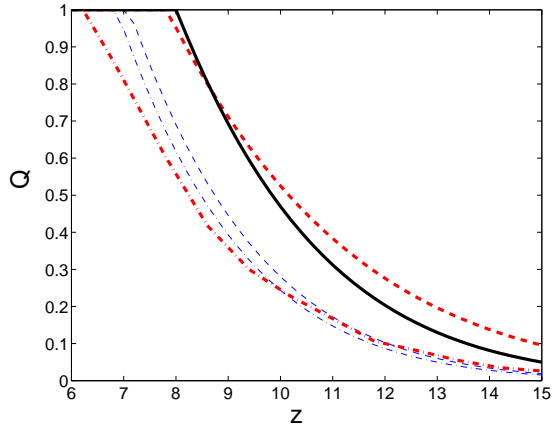} \caption{Comparison of
reionization histories with $\zeta = 12 $ between cosmologies that
have different mass functions and cosmological parameters.  The thick
solid is the standard cosmology adopted in this paper with the
Press-Schechter mass function. Other curves show the same model except
for $\sigma_8 = 0.8$ (thin dot-dashed) and for tilt $n = 0.95$ (thin
dashed).  We also plot the $Q(z)$ for the Sheth-Tormen mass function
(thick dashed) and Press-Schechter with a top hat real space filter
calculated from random walks (thick dot-dashed).}
\label{collfig}
\end{figure}

Even though the least well-known parameter in our model is $\zeta$,
there is enough uncertainty in the cosmological parameters and the
mass function to strongly affect the morphology of reionization.
Beyond the usual uncertainty in $\Omega_m$, $n$, etc., there is also
the shape and amplitude of the mass function at high redshifts; each
choice will yield a different reionization history.  Simulations are
our best avenue for ascertaining the correct mass function.  However,
to date most high-resolution simulations have only investigated epochs
well after reionization.  We found two studies of the mass function at
$z \ga 10$.  \citet{jang01} showed that the Press-Schechter mass
function agrees reasonably well with the mass function from their
simulation at $z = 10$.  However, their $1 h^{-1}$ comoving Mpc cube
only allowed them to probe the low-mass end.  The other was
\citet{reed03}, which concludes that the \citet{sheth02} mass function
overpredicts substantially the mass function in simulations for $z >
10$.  In Figure \ref{collfig}, we plot the expected
reionization history $Q(z)$ for several mass functions: Sheth-Tormen
(thick dashed), the usual Press-Schechter function (thick solid), and
Press-Schechter with a top hat real space filter computed from random
walks (thick dot-dashed).  Surprisingly, the top hat real space filter
gives a significantly smaller collapse fraction than the sharp
$k$-space filter implicit in the extended Press-Schechter formalism
(by about a factor of $0.5$).  This is because the variance in the
density field within the top hat window function is smaller than the
variance within the sharp $k$-space window function.  Since the
barrier $\delta_c = 1.686$ is derived from top hat considerations, it
could well be that the top hat in real space gives a more realistic
mass function at these high redshifts.

Several observations of large scale structure predict a lower value of
$\sigma_8$ than the $\sigma_8 = 0.9$ from CMB data
\citep{tegmark04_SDSS}.  We plot $Q(z)$ for $\sigma_8 = 0.8$ (but with
other parameters fixed) in Figure \ref{collfig} (thin dot-dashed).
This value for $\sigma_8$ causes reionization to end later.  In
addition, CMB observations are consistent with the tilt $n$ in the
primordial power spectrum being slightly less than the
Harrison-Zel'dovich choice $n = 1$ \citep{spergel03, tegmark04_SDSS}.
Even a slightly smaller tilt, $n = 0.95$ (thin dashed), causes the
universe to reionize significantly later.

Neither these different cosmological parameters nor a different mass
function significantly alters the duration of reionization.  Because
the kSZ amplitude is most affected by the duration of reionization,
all of the reionization histories in Figure \ref{collfig} should
produce similar patchy signals. However, the amplitude of the tSZ
effect scales as $\sim \sigma_8^{6-9}$ \citep{seljak01} and the OV 
signal scales as
$\sim \sigma_8^{4-6}$ \citep{zhang04}.  Thus, if $\sigma_8$ is smaller
than $0.9$ these anisotropies will be suppressed, and the patchy
signal will be more prominent.  

If the dark matter is warm, this will have an even more drastic effect
on the ionized fraction \citep{barkana01b, kitano05}.  Warm dark
matter will create a cutoff in the transfer function at small scales
such that the collapsed fraction is significantly smaller.  While warm
dark matter cannot make universes with a high optical depth
\citep{yo03b}, this can be supplemented by having a
decaying particle ionize the universe at higher redshifts
\citep{hansen04}.

Finally, we note that because the OV effect is dominated by lower
redshift structure formation, the uncertainty in the high-$z$ mass
function has only a negligible effect on that signal.

\section{Discussion}
\label{discussion}

We find that there is a clear and measurable difference between the
amplitude of the kSZ signal for models where Population II stars
dominate reionization, where Population III stars are also important
(or reionization is extended for some other reason), and where there
are a large number of hard photons (or uniform reionization).  At
scales where the kSZ effect dominates over other primordial anisotropies, the
amplitude of the patchy signal differs by as much as $10 {\mu {\rm K}}^2$.
Other aspects of the patchy signal also differ between reionization
histories, such as the multipole ${\it l}$ where the patchy power
spectrum peaks.  The peak of the power spectrum is primarily
determined by the distribution of bubble sizes when $Q \sim 0.5$ or,
for our extended reionization scenarios, by the bubble size when the
ionized fraction ``freezes.''  Ideally, one could fit a simple (but
physically motivated) two or three parameter model to the
observations.  Of course, reionization is a very complex process and
many parameters are needed to describe its morphology in detail (in
our model, these are essentially the parameters determining $\zeta$ as
a function of redshift).  To add to the difficulty, observations cannot
separate the OV and patchy contributions to the CMB power spectrum,
requiring simultaneous fits to both.\footnote{Another ``contaminant'' of
the power spectrum at these scales is the ``moving clusters of
galaxies" effect \citep{molnar00}.  Its amplitude on the scales of
interest is smaller than that of the kSZ signal (${\it l}^2 C_l/{2
\pi} \sim 10^{-14}-10^{-13}$) though not obviously negligible.}  The
OV power is dominated by post-reionization sources and contains much
less information about the morphology of reionization than the patchy
signal.

The OV contamination of the patchy signal might not be a
problem. First, because a substantial fraction of the OV power comes
from low redshift objects, much of it may be removable from the
signal.  Also, future 21cm observations will help separate these two
contributions (e.g., FZH04, \citealt{furlanetto04b,salvaterra05,zal04}).
Furthermore, with a fixed cosmology, the OV contribution to the power
spectrum is determined only by the distribution of gas.  To zeroth
order, we know that gas traces the dark matter distribution, which can
be successfully modeled with halo theory.  At large overdensities our
understanding of the gas distribution is incomplete.  But for ${\it l}
< 10^4$, the OV power is relatively independent of the gas
distribution under reasonable assumptions.  This is evident in Figure
\ref{OVfig}, where the thin black curve is the OV signal in the most
extreme case in which the baryons trace the dark matter.
\cite{zhang04} contend that one can reasonably calculate the OV
contribution without any free parameters as long as the evolution of
the ionized fraction is known.  The ionized fraction at each redshift
is a byproduct of the model for patchy reionization.  Therefore, the
OV signal -- which has a much flatter power spectrum -- is predicted
by our model with no additional parameters (to lowest order) and could
help break the degeneracy between the redshift of overlap and other
parameters.  Of course, minor differences do remain in the calculation
of the OV signal between different computational approaches (see \S
\ref{kSZ}).  We expect that future research, particularly simulations
incorporating realistic gas heating, will help to resolve these small
discrepancies.

To this point, we have concentrated on the information contained in
the kSZ signal.  However, it will also contaminate fits to the
primordial anisotropies.  \citet{knox98}, \citet{santos03} and
\citet{zahn05} investigated how the signal from patchy reionization
will bias cosmological parameter determination for \emph{WMAP} and
\emph{Planck}.  While the net effect is insignificant for \emph{WMAP},
it will be quite substantial for \emph{Planck}, which can measure
$C_{\it l}$ to ${\it l} = 2500$.  In this case, the bias for many
cosmological parameters is comparable to the 1-$\sigma$ statistical
error.  \citet{santos03} propose adding an additional parameter, the
effective amplitude of the kSZ power spectrum, in future fits to
\emph{Planck} data, which is justified by the fact that the kSZ signal
is fairly flat in the regime where it affects the fit.  A similar
conclusion holds for our model for reionization \citep{zahn05}.
The polarization-polarization CMB power spectrum might
be more reliable for measuring the cosmological parameters at
${\it l} \gtrsim 1000$ since it is nearly unaffected by patchiness
\citep{zahn05}.

Other probes of the reionization epoch may help to restrict the set of
possible reionization histories, facilitating the interpretation of
the kSZ signal measured by ACT or SPT.  A more precise measurement of
the optical depth $\tau$ from measurements of large-scale CMB
polarization will significantly reduce the set of viable models.  In
addition, if future observations of high-redshift quasars or 21 cm
emission from neutral hydrogen can pin down the redshift of overlap,
this will again reduce the set of plausible models.  To the extent
that the current large optical depth measurement implies a long period
of extended reionization, the patchy contribution should extend over a
long redshift interval and have a substantial amplitude.  This will
allow future experiments to distinguish reionization by stellar
sources, hard photons, and decaying particles, and alleviate many of
the degeneracies from large-scale CMB polarization data alone.

\acknowledgments This work was supported in part by NSF grants AST
02-06299, AST-0098606 and AST 03-07690, by NASA ATP grants
NAG5-12140, NAG5-13292, and NAG5-13381, and by the David \& Lucille
Packard Foundation and the Sloan Foundation.

\begin{appendix}

\section{The Baryon Distribution}
\label{pbb}

We construct $P_{\delta_b \delta_b}$ from the halo model by using NFW
dark matter halo profiles with the Jenkins mass function and by
setting $P_{\delta_b \delta_b}(k) = F(k/k_F)^2 \,P_{\delta
\delta}(k)$, where $F(x)$ filters large $k$ modes in order to
approximate the effects of finite gas pressure.  The Jeans length is
often not the appropriate filtering scale $k_F$.  \citet{gnedin98a}
and \citet{gnedin98b} suggest $k_F(z) = 34 \,\sqrt{\Omega_m(z)} h$
Mpc$^{-1}$ for thermal history-dependent filtering and
\begin{equation}
	F(x) = \frac{1}{2} \left[ e^{-x^2} + \frac{1}{{(1 + 4
	x^2)}^{1/4}} \right].
\end{equation}
They find that for $\delta < 10$, this filtering function produces
errors smaller than $10\%$ in the gas power spectrum $P_{\delta_b
\delta_b}$.  We use this scheme for $z < 8$ and set $F(x) = 1$ for $z
> 8$, where the gas distribution is less well-understood and it is
thought that cooling is more efficient.  Note that the difference in
the kSZ signal between reasonable filtering schemes for high redshifts
is minute on the scales of interest.

\section{Configuration Space Calculation of the kSZ Effect}
\label{realspace}

For the patchy terms, namely $\langle x_i \, x_i' \rangle \langle v \,
v'\rangle $ and $\langle x_i \, \delta_b'\rangle \langle v \,
v'\rangle $, we perform the kSZ calculation in configuration space
using equation (\ref{wtheta}).  We first use linear theory to
construct the velocity field.  In that case, the velocity correlation
function is
\begin{equation}
\langle v(\eta, {\bf \hat{n}})\, v(\eta',
{\bf \hat{n}'})\rangle = [\xi_{\perp}(r)] \, {\bf \hat{n}} \cdot {\bf
\hat{n}'} + [\xi_{||}(r) - \xi_{\perp}(r)] \, \frac{({\bf r}
\cdot {\bf \hat{n}})({\bf r} \cdot {\bf \hat{n}'})}{r^2},
\end{equation}
where $v({\bf \hat{n}}) = {\bf v}({\bf \hat{n}}) \cdot {\bf \hat{n}}$,
${\bf r} \equiv {\bf x} - {\bf x'}$ and $\xi_{||}$ and $\xi_{\perp}$
are the autocorrelation functions of the velocity parallel and
perpendicular to ${\bf r}$:
\begin{equation}
\binom{ \xi_{||}}{\xi_{\perp}}= - {\it f}^2 H_0^2 \int_0^{\infty}
\frac{dk}{2 \pi^2 k} P_{\rm lin}(k) \binom{ j_0'(k x)/x}{k ~ j_0''(k x)}.
\label{vv}
\end{equation}
Here primes denote derivatives with respect to $kx$.  If we put the
correlation functions constructed here and in \S \ref{ps} into
equation (\ref{wtheta}), we can calculate $w(\theta)$ directly.  For
${\it l} \gg 1$, the angular power spectrum is then
\begin{equation}
    C_{\it l} \approx 2 \pi \int_0^{\infty} \theta d\theta \, J_0({\it l}
    \theta)\, w(\theta).
\end{equation}

\section{Derivation of $Q(z)$ and Mass Function for a Linear Barrier}
\label{massfunc}

Our aim in this section is to derive (or rederive) some relations we
will need for a linear barrier using the extended Press-Schechter
formalism \citep{bond91, lacey}.  The advantage of this approach is
that the probability distribution of $\delta_k$ -- the real-space
density within a $k$-space top hat filter of radius $k$ -- becomes the
solution to a diffusion equation.  This diffusion equation can then be
solved for various boundary conditions (corresponding to the specified
barrier).  We first rederive the familiar result for a constant
barrier adopting the notation of \citet{scannapieco02} and then expand
the approach to the case of a linear barrier.  The variance today in a
region defined by a sharp $k$-space window function is
\begin{equation}
S_k = \sigma^2(k) = \frac{1}{2 \pi^2} \int_0^k dk k^2 P_{\rm
lin}(k), \label{variance}
\end{equation}
\noindent where $P_{\rm lin}(k)$ is the linear power spectrum at $z =
0$.  As we change the smoothing scale -- or increase $k$ in equation
(\ref{variance}) -- each step $\Delta S_k$ is uncorrelated with the
previous $S_k$ (this is not true for other filter choices).
Therefore, we can write a simple evolution equation for $Q(\delta, S_k)$,
the probability distribution for the density field at smoothing scale
$k$ \citep{bond91}
\begin{equation}
Q(\delta, S_k) = \frac{1}{\sqrt{2 \pi \,{\Delta S_k}}}
\int_{-\infty}^{\infty} {d\Delta\delta
 \,{\rm exp}\left[ -\frac{(\Delta\delta)^2}{2\,
{\Delta S_k}} \right]\,Q(\delta - \Delta\delta, S_k - \Delta S_k)}.
\label{Qeqn}
\end{equation}
If we expand $Q$ to second order in $\Delta \delta$ and perform
the $\Delta \delta$ integral, equation (\ref{Qeqn}) reduces to
\begin{equation}
Q(\delta, S_k) = Q(\delta, S_k-\Delta S_k) + \frac{1}{2}
\frac{\partial^2 Q(\delta, S_k-\Delta S_k)}{\partial \delta^2}\,
\Delta S_k,
\end{equation}
noting that $\langle \Delta \delta \rangle = 0$ and that $\langle
(\Delta \delta)^2 \rangle = \Delta S_k$.  Rearranging the above
equation and keeping terms to linear order in $\Delta S_k$ gives a
diffusion equation
\begin{equation}
 \frac{\partial Q(\delta,S_k)}{\partial S_k} = \frac{1}{2}
 \frac{\partial^2 Q(\delta, S_k)}{\partial
 \delta^2}. \label{diffusion} 
\end{equation}
\noindent In the absence of any barrier, the solution to this
equation is
\begin{equation}
Q_0(\delta, S_k) = \frac{1}{\sqrt{2 \pi \, S_k}}{\rm
exp}\left[-\frac{\delta^2}{2\, {S_k}}\right]. \label{Q0}
\end{equation}
The solution to a diffusion equation is uniquely specified by the
initial and boundary conditions.  Thus, for a constant barrier $B(k) =
B_0$, we can guess the unique solution that satisfies the boundary
condition $Q(B_0, S) = 0$, namely $Q = Q_0(\delta, S_k) - Q_0(2 B_0
-\delta, S_k)$.  The second term is an ``image'' that is placed to
cancel the contribution of the source term on the boundary
\citep{chan43,bond91}.

For the bubble problem, we are ultimately interested in a solution to
this diffusion equation for a barrier that is linear in the variance
$B(k) = B_0 + B_1 S_k$. This amounts to the boundary condition
$Q(B(k), S_k) = 0$ for the PDE (\ref{diffusion}). To find a solution,
let us rescale the variable $\delta$ to $y = B_1(\delta - B_1 S_k)$
such that the diffusion equation transforms to
\begin{equation}
 \frac{\partial Q}{\partial S_k}= \frac{B_1^2}{2}
\left[\frac{\partial^2 Q}{\partial y^2} + 2 \, \frac{\partial
Q}{\partial y}\right], \label{lindiff}
\end{equation}
and the boundary condition becomes $Q(B_1 \,B_0, S_k) = 0$. Unlike
the constant barrier problem, the image method is not a fruitful
way to solve this problem.  Instead, let us look for solutions of
the form $Q(y, S_k) = g(y)\,f(S_k)$.  Then equation
(\ref{lindiff}) becomes
\begin{equation}
\frac{1}{f}\frac{\partial f}{\partial S_k} = \frac{B_1^2}{2
\,g} \left[ \frac{\partial^2 g}{\partial y^2} + 2\,\frac{\partial
g}{\partial y} \right] =  \lambda.
\end{equation}
\noindent Dropping constant factors, the general solutions for $f$
and $g$ are $f(S_k) = {\rm exp}(\lambda \,S_k)$ and $g(y) = {\rm
exp}[(-1 \pm {\it i} X) \, y]$ where $X = - {\it i} \sqrt{1 +2
\lambda/B_1^2}$. If we drop the eigenfunctions which do not
satisfy the boundary conditions, our solution $Q(y, S_k)$ becomes
\begin{equation}
Q_{\rm lb}(y, S_k) = \int_{0}^{\infty}{dX} h(X) ~{\rm sin}[X (y - B_0
B_1)]~\exp \left[-y-\frac{B_1^2}{2}(1 + X^2)S_k \right]. \label{solwh}
\end{equation}
To determine $h(X)$, we impose the initial condition.  In the
$(\delta, S)$ coordinates, the initial condition is $ \delta_D(\delta)
\equiv {\rm lim}_{S_k \rightarrow 0}\,{\rm exp}[- \delta^2/(2\,
{S_k})]/\sqrt{2 \pi \, S_k}$.  If we substitute $y$ for $\delta$ this
becomes ${\rm lim}_{S_k \rightarrow 0}\, {\rm exp}[{-B_1^2 \, S_k/2 -y
- y^2/(2 B_1^2 S_k)}]/ \sqrt{2 \pi S_k} = B_1 \, \delta_D(y)\, {\rm
exp}(-y)$.  Using this, we find that $h(X) = B_1/\pi~ {\rm sin}(X B_0
B_1)$, and we can now integrate equation (\ref{solwh}):
\begin{eqnarray}
Q_{\rm lb}(y, S_k) & =  &\frac{1}{\sqrt{2 \pi S_k}} \, 
{\rm exp}{[ -\frac{B_1^2 S_k}{2} -y ]}
\left\{ \exp \left[-\frac{y^2}{2 B_1^2 S_k}\right] - \exp \left[
  -\frac{(y- 2B_0 B_1)^2}{2B_1^2 S_k} \right] \right\} \label{imagesoln1}\\
& = &\frac{1}{\sqrt{2 \pi S_k}}
\left\{ \exp \left[ -\frac{\delta^2}{2 \,S_k} \right] -
\exp \left[ -\frac{|2(B_0 + {\it i}\sqrt{B_0 B_1 S_k}) - \delta|^2}{2
    S_k} \right] \right\}.
    \label{imagesoln2}
\end{eqnarray}
\noindent The solution
(\ref{imagesoln1}) vanishes at the boundary $y = B_0 B_1$, as
required. In addition, equation (\ref{imagesoln2}) shows that there is
an analogous form to the familiar ``image'' decomposition for the
constant barrier problem.

To  obtain the  ionized fraction, we integrate over  all possible
$\delta$  at  $S_k  =   \sigma_{\rm  min}^2$,  which  yields  equation
(\ref{onepoint_text}).  We can obtain  the mass function as well.  The
probability  that a  trajectory crosses  the barrier  in  the interval
$S_k$ to $S_k + dS_k$ is
\begin{equation}
P(S_k) = -\frac{d}{d S_k}\int_{-\infty}^{B_0 B_1} \left(
\frac{dy}{B_1} \right )~ Q_{\rm lb} = -\left[ \frac{B_1}{2}\frac{\partial
Q_{\rm lb}}{\partial y} \right]_{-\infty}^{B_0 B_1} = \frac{B_0}{\sqrt{2 \pi \,
S_k^3}} {\rm exp} \left[-\frac{B(k)^2}{2\, {S_k}} \right],
\label{eq:psk}
\end{equation}
where the second inequality uses the diffusion equation and the
boundary condition $Q_{\rm lb}(B_0 B_1, S_k) = 0$.  Equation
(\ref{eq:psk}) agrees with the results of \citet{sheth98}, which were
derived through a different method. The number density of bubbles of
mass $m$ is then $n(m) = \rho_0/m \, P(S_k) \, |{\deriv S_k}/{\deriv m}|$, which yields equation (\ref{eq:dndm}).

\end{appendix}

\bibliographystyle{apj}

\end{document}